\documentclass[12pt]{article}
\usepackage[top=1in, bottom=1in, left=1.25in, right=1.25in]{geometry}
\usepackage{float}
\usepackage{color}
\usepackage{tikz}
\usepackage{setspace}
\usepackage[toc,page]{appendix}
\usepackage{amssymb}
\usepackage{showlabels}
\usepackage[normalem]{ulem}
\usepackage{multirow}
\usepackage{tikz}
\usepackage[toc]{appendix}
\usepackage{chngcntr}
\usetikzlibrary{shapes,arrows}
\usepackage{subfigure}
\usepackage[round]{natbib}
\usepackage{latexsym}
\usepackage{amsmath}
\usepackage{graphicx}
\usepackage{caption}
\usepackage{soul}
\providecommand{\keywords}[1]{\textbf{\textit{Key words:}} #1}
\usepackage{bm}
\usepackage[pdftex,hypertexnames=false,linktocpage=true]{hyperref}
\hypersetup{colorlinks=true,linkcolor=blue,anchorcolor=blue,citecolor=blue,filecolor=blue,urlcolor=blue,bookmarksnumbered=true,pdfview=FitB}

\newcommand{\biblist}{\begin{list}{}
		{\listparindent 0.0cm \leftmargin 0.50cm \itemindent -0.50 cm
			\labelwidth 0 cm \labelsep 0.50 cm
			\usecounter{list}}\clubpanelty4000\widowpanelty4000}
	\newcommand{\ebiblist}{\end{list}}

\def\T{{ \mathrm{\scriptscriptstyle T} }}

\def\0{{\bf 0}}

\def\mis{{\rm mis}}
\def\obs{{\rm obs}}
\def\pr{{\text{pr}}}
\def\FULL{\textsc{FULL}}
\def\TRUE{\textsc{TRUE}}
\def\BMA{\textsc{BMA}}

\def\LASSO{\textsc{LASSO}}
\def\var{{\text{var}}}
\def\MI{\textsc{MI}}

\newtheorem{theorem}{Theorem}
\newtheorem{lemma}{Lemma}

\newtheorem{remark}{Remark}

\newtheorem{assumption}{Assumption}
\def\boxit#1{\vbox{\hrule\hbox{\vrule\kern6pt\vbox{\kern6pt#1\kern6pt}\kern6pt\vrule}\hrule}}

\title{\LARGE\bf Accounting for model uncertainty in multiple imputation under complex sampling}
\author{Gyuhyeong Goh\thanks{Department of Statistics, Kansas State University, Manhattan, KS 66506, U.S.A.} \and Jae Kwang Kim \thanks{Department of Statistics, Iowa State University, Ames, IA 50011, U.S.A.}}

\begin{document}
\baselineskip .3in
\maketitle

\begin{abstract}
Multiple imputation provides an effective way to handle missing data. When several possible models are under consideration for the data, the multiple imputation is typically performed under a single-best model selected from the candidate models. This single model selection approach ignores the uncertainty associated with the model selection and so leads to underestimation of the variance of multiple imputation estimator. In this paper, we propose a new multiple imputation procedure incorporating model uncertainty in the final inference. The proposed method incorporates possible candidate models for the data into the imputation procedure using the idea of Bayesian Model Averaging (BMA). The proposed method is directly applicable to handling item nonresponse in survey sampling.  
Asymptotic properties of the proposed method are investigated. A limited simulation study confirms that our model averaging approach provides better estimation performance than the single model selection approach.
\end{abstract}

\keywords{Approximate Bayesian Computation, Bayesian Model Averaging, Informative sampling, Item nonresponse.}

\newpage
\section{Introduction}\label{sec:1}
 In survey sampling, item nonresponse is frequently encountered and imputation is a popular technique for handling item nonresponse. Multiple imputation has been proposed by \cite{Rubin:1978,Rubin:1987} as a  general tool for accessing the uncertainty  of sample estimates in the presence of imputed values. See \cite{Little:2002} for a comprehensive overview of the multiple imputation methods.

 While multiple imputation has been promoted in many application areas, the theory for multiple imputation is somewhat limited. \cite{Schenker-Welsh:1988}, \cite{Wang-Robins:1998}, \cite{Nielsen:2003} and \cite{Kim-Yang:2017} develop rigorous asymptotic theories  for multiple imputation estimator of the parameters in the specified model. \cite{Meng:1994}, \cite{Kim:2006}, \cite{Yang-Kim:2016} and \cite{Xie-Meng:2017} discuss issues associated with the self-efficiency assumption for multiple imputation. However, all the above mentioned papers are developed under the correctly specified model. In practice, the true model is unknown and we often  use a model selection procedure  in addition to parameter estimation to implement multiple imputation. How to incorporate the model selection uncertainty into multiple imputation is an important research gap that has not been addressed  in the literature.


  In this paper, we propose a new multiple imputation procedure that covers model selection and parameter estimation simultaneously. Specifically, we incorporate  possible candidate models for the data into the imputation procedure using the idea of Bayesian Model Averaging \citep{Madigan:1994, Hoeting:1999}. The proposed method is justified from a rigorous asymptotic theory and also confirmed in a limited simulation study. By including the model selection uncertainty into the imputation procedure, we can obtain valid inference  with multiple imputation. {The proposed method can be particularly useful in variable selection problem in the regression models} \citep{Freedman:1983, Tibshirani:1996, Raftery:1997, Efron:2004}, as illustrated in Section 5. 

   The paper is organized as follows. In Section 2, the basic setup is introduced and the classical multiple imputation procedure is reviewed. In Section 3, the new multiple imputation method is  proposed. In Section 4, some asymptotic properties of the proposed method is established. In Section 5, results from a limited simulation study are presented to confirm the validity of the proposed method. Some concluding remarks are made in Section 6.

\section{Basic setup}\label{sec:2}
Suppose that the finite population $\{(x_i,y_i): i=1,\ldots,N\}$ is a random sample of size $N$ from an infinite population $\zeta$ with joint density $f(y \mid x)f(x)$, where $y$ is a scalar response variable and $x$ is a vector of explanatory variables. We assume that $f(y\mid x)=f(y\mid x;\theta)$ is a parametric model with parameter $\theta\in \Theta$, whereas $f(x)$ is completely unspecified. Let  $A$  be the set of sample indices  obtained from the finite population under a probability sampling design. In addition, we consider the setting in which $y_i$ are subject to missingness, while $x_i$ are fully observed. We denote by $D_n=\{X_n,Y_n\}$ the complete data, where $X_n=\{x_i:i\in A \}$ and $Y_n=\{y_i:i\in A \}$.  Let $Y_{\obs}$ and $Y_{\mis}$ be the observed part and the missing part of $Y_n$, respectively. Thus, we can denote  the observed data by $D_{\obs}=\{X_n,Y_\obs\}$. Define $\delta_i\in\{0,1\}$ to indicate the observed response status; that is, $\delta_i=1$ if $y_i$ is observed and $\delta_i=0$ otherwise. We assume that the missing mechanism is missing at random (MAR) at the population level \citep{berg2015}
in the sense that
\begin{eqnarray*}\label{PMAR}
\pr(y_i\in B \mid x_i,\delta_i=1)=\text{pr}(y_i\in B \mid x_i)
\end{eqnarray*}
for any measurable set $B$ and for all $x_i$. Given the population model $f(y\mid x;\theta)$, we further assume that certain restrictions on the parameter $\theta$ lead to the parsimonious true data-generating model. In regression analysis, for example, if some explanatory variables are irrelevant to the response variable, then the true model is obtained by setting the corresponding coefficients to zero.

Let $\kappa$ be the true model and $\theta_\kappa$ be the active parameter under model $\kappa$. We denote the true population model by $f(y \mid x;\theta_\kappa,\kappa)$, which can be obtained by imposing the parameter restriction of model $\kappa$ on $f(y \mid x;\theta)$. If the data are fully observed, a consistent estimator of $\theta_\kappa$ under model $\kappa$, say $\hat{\theta}_\kappa$, can be easily obtained by maximizing the pseudo log-likelihood function
\begin{eqnarray}\label{pMLE}
l_w(\theta_\kappa)=\sum_{i\in A} w_i \log f(y_i \mid  x_i;  \theta_\kappa,\kappa),
\end{eqnarray}
where $w_i$ is the sampling weight for unit $i$. Use of the  maximum pseudo likelihood estimator is well studied in the literature \citep{Binder:1983, Godambe:1986}. Furthermore, under the regularity conditions of \citet{Fuller:2009}, the asymptotic normality holds for the maximum pseudo likelihood estimator $\hat{\theta}_\kappa$ as follows:
\begin{eqnarray*}\label{CLT:1}
\left. \left\{\hat{V}(\hat{\theta}_\kappa)\right\}^{-1/2}\left(\hat{\theta}_\kappa-\theta_\kappa \right)\right| ( \theta_\kappa,\kappa) \stackrel{\mathcal{L}}{ \longrightarrow  } \mathcal{N}(0,I),
\end{eqnarray*}
as $n\to \infty$, where $\stackrel{\mathcal{L}}{ \longrightarrow  }$ denotes the convergence in distribution,
 $\hat{V}(\hat{\theta}_\kappa)$ is a consistent estimator of $\var(\hat{\theta}_\kappa\mid \theta_\kappa,\kappa)$ and $I$ is the identity matrix. Using the Sandwich formula \citep{Binder:1983}, for example, a consistent variance estimator can be obtained by
 \begin{eqnarray}\label{var:est}
\hat{V}(\hat{\theta}_\kappa)=\left\{ \sum_{i\in A} w_i l''_i( \hat{\theta}_\kappa)  \right\}^{-1}\left\{ \sum_{i\in A} \sum_{j \in A} \frac{\pi_{ij}-\pi_i\pi_j}{\pi_{ij}}w_i l'_i( \hat{\theta}_\kappa)  w_j l'_j( \hat{\theta}_\kappa)^\T \right\}\left\{ \sum_{i\in A} w_i l''_i( \hat{\theta}_\kappa)^\T  \right\}^{-1},
\end{eqnarray}
 where $l'_i(\theta_\kappa)= \partial \log f(y_i \mid  x_i;  \theta_\kappa,\kappa)/\partial \theta_\kappa $, $l''_i(\theta_\kappa)= \partial^2 \log f(y_i \mid  x_i;  \theta_\kappa,\kappa)/(\partial \theta_\kappa \partial \theta_\kappa^\T) $, $\pi_i$ is the inclusion probability for element $i$ and $\pi_{ij}$ is the joint inclusion probability for elements $i$ and $j$.

  However, in the presence of missing values, our challenge is that the pseudo likelihood estimating formula in \eqref{pMLE} is not directly applicable.
To handle missing data, imputation, the technique of filling in missing values in the data, can be  considered. The major attractive feature of imputation is that the complete-data estimation technique can be directly applied to an imputed data set.  Multiple imputation, proposed by \citet{Rubin:1987},  has been increasingly popular to incorporate the uncertainty associated with predicting missing values into the imputation estimator. In our setup, given the correct specification of the true model $\kappa$,  the multiple imputation procedure can be summarized as the following three steps:
\begin{enumerate}
\item[1.] (Imputation) For $m=1,\ldots, M$, generate $Y^{(m)}_{\mis}$ from $p(Y_{\mis}\mid D_{\obs},\kappa)$ independently, where $p(Y_{\mis}\mid D_{\obs},\kappa)$ is the conditional distribution of the missing response $Y_{\mis}$ given the observed data $D_{\obs}$ and model $\kappa$. Using the imputed values, we can create $M$ complete data sets, $D_n^{(1)}, \ldots, D_n^{(M)}$, where $D_n^{(m)}=\{X_n,Y_n^{(m)}\}$ and $Y_n^{(m)}=\{Y_\obs,Y_\mis^{(m)}\}$.
\item[2.] (Analysis) For each of the complete data sets, compute the  maximum pseudo likelihood estimator of $\theta_\kappa$ and its variance estimator, $\hat{\theta}_\kappa^{(m)}$ and $\hat{V}(\hat{\theta}_\kappa^{(m)})$, from \eqref{pMLE} and \eqref{var:est}, respectively.
\item[3.] (Pooling) Applying \citet{Rubin:1987}'s formula, the multiple imputation estimator and its variance estimator are obtained as
\begin{eqnarray*}\label{rubin:eq}
\hat{\theta}_{\MI,\kappa}=M^{-1}\sum_{m=1}^M\hat{\theta}_\kappa^{(m)},\quad \hat{V}_{\MI,\kappa}=W_{\kappa}+\left(1+M^{-1}\right)B_{\kappa},
\end{eqnarray*}
where $W_{\kappa}=M^{-1}\sum_{m=1}^M\hat{V}(\hat{\theta}_\kappa^{(m)})$ and $B_{\kappa}=(M-1)^{-1}\sum_{m=1}^M(\hat{\theta}_\kappa^{(m)}-\hat{\theta}_{\MI,\kappa})(\hat{\theta}_\kappa^{(m)}-\hat{\theta}_{\MI,\kappa})^\T$.
\end{enumerate}

In the multiple imputation procedure, the imputation step is commonly implemented through the data augmentation algorithm of \citet{Tanner:1987}, which iteratively generates values of $\theta_\kappa$ and $Y_{\mis}$ from $p(\theta_\kappa\mid D_n,\kappa)$ and $p(Y_{\mis}\mid D_{\obs},\theta_\kappa,\kappa)$ under the specified model $\kappa$.

When the sampling design is informative \citep{Pfeffermann:1993}, implementing the data augmentation algorithm is not straightforward because the sample-data likelihood $p(Y_n\mid X_n,\theta_\kappa,\kappa)$ is unknown under informative sampling,
 hence the posterior distribution, $p(\theta_\kappa\mid D_n,\kappa)\propto p(Y_n\mid X_n,\theta_\kappa,\kappa)p(\theta_\kappa \mid \kappa)$, is also unknown, where $p(\theta_\kappa \mid \kappa)$ denotes the prior distribution of $\theta_\kappa$ under model $\kappa$. As an alternative, \citet{Kim-Yang:2017} recently proposed a new data augmentation algorithm that can implement multiple imputation under informative sampling.  In the proposed method of \citet{Kim-Yang:2017}, the sample-data likelihood is replaced by the sampling distribution of the maximum pseudo likelihood estimator and  multiple imputation can be used to handle  missing data even under informative sampling.

 However, existing imputation methods is only applicable to the case that the model $\kappa$ is known or fixed. In practice, the true model is mostly unknown. Thus, it is common to select a single best model, say $\hat{\kappa}$, from a set of candidate models by model selection techniques and then to apply the multiple imputation procedure after replacing $\kappa$ with $\hat{\kappa}$. Similarly to the problem of single imputation, the most severe limitation of the single model selection approach is that the true model is unknown and yet the selected model $\hat{\kappa}$ is treated as the true model. As noted by \citet{Hoeting:1999},   ignoring the model uncertainty in model selection can lead to underestimation of uncertainty about the parameter of interest. To address this issue, in the following section, we introduce a new multiple imputation procedure which incorporates the model uncertainty into the final inference.

\section{Proposed method}\label{sec:3}

From a Bayesian perspective, for given the observed data $D_{\obs}$, the unknown true model $\kappa$ is treated as a random variable with the posterior model probability $p(\kappa\mid D_{\obs})$, see \citet{Gelfand:1994, Kass:1995, Hoeting:1999} and references therein. A complete Bayesian solution to the model selection problem, often referred to as Bayesian model averaging, is to average over all possible models weighted by their posterior model probabilities \citep{Raftery:1997}. In a similar spirit, we propose a new multiple imputation estimator which is an weighted average of the multiple imputation estimators over candidate models under consideration. To this end, we first decompose the full parameter $\theta$ into two parts, $\theta_\kappa$ and $\theta_{\bar{\kappa}}$, where $\theta_{\kappa}$ and $\theta_{\bar{\kappa}}$ correspond to the active parameter and the restricted parameter under model $\kappa$, respectively. For the sake of notational simplicity, without loss of generality, we assume that $\theta=(\theta_\kappa^\T,\theta_{\bar{\kappa}}^{\T})^\T$. Let $|\kappa|$ and $|\bar{\kappa}|$ be the numbers of active elements and restricted elements in $\theta$ under model $\kappa$, respectively. For model $\kappa$, we assume that $\theta_{\bar{\kappa}}=\theta^0_{\bar{\kappa}}$ , where $\theta^0_{\bar{\kappa}}$ is a constant vector. In the context of variable selection, for example, we set $\theta_{\bar{\kappa}}=0$ so that the inactive explanatory variables are excluded in the restricted model $f(y\mid x;\theta_\kappa,\kappa)$. We now introduce a new multiple imputation procedure as follows:
\begin{enumerate}
\item[1$^\star$.] (Model sampling) Generate $\kappa^{(m)}$ from $p(\kappa \mid D_{\obs})$ independently for $m=1,\ldots, M$.
\item[2$^\star$.] (Imputation) Under the generated model $\kappa^{(m)}$, draw $Y^{(m)}_{\mis}$ from
$p(Y_{\mis} \mid D_{\obs},\kappa^{(m)})$ independently for $m=1,\ldots, M$. Then, define $M$ imputed data sets by $D_n^{(1)}, \ldots, D_n^{(M)}$, where $D_n^{(m)}=\{X_n,Y_n^{(m)}\}$ and $Y_n^{(m)}=\{Y_\obs,Y_\mis^{(m)}\}$.
\item[3$^\star$.] (Analysis) Using each pair of the complete data sets $D_n^{(m)}$ and the generated model $\kappa^{(m)}$, compute $\hat{\theta}^{(m)}$ and $\hat{V}^{(m)}$ using (\ref{pMLE}) and (\ref{var:est}) under the constraint on $\theta_{\bar{\kappa}^{(m)}}$.
\item[4$^\star$.] (Pooling) The multiple imputation estimator and its variance estimator are computed as
\begin{eqnarray}\label{new:rubin}
\hat{\theta}_{\MI}=M^{-1}\sum_{m=1}^M\hat{\theta}^{(m)},\quad \hat{V}_{\MI}=W+\left(1+M^{-1}\right)B,
\end{eqnarray}
where $W=M^{-1}\sum_{m=1}^M\hat{V}^{(m)}$ and $B=(M-1)^{-1}\sum_{m=1}^M(\hat{\theta}^{(m)}-\hat{\theta}_{\MI})(\hat{\theta}^{(m)}-\hat{\theta}_{\MI})^\T$.
\end{enumerate}
In the newly-developed multiple imputation procedure, to implement the first two steps 1$^\star$ and 2$^\star$ (Model sampling and Imputation) simultaneously, we can use a variation of the data augmentation algorithm \citep{York:1995} by iterating the following three steps until convergence:
\begin{enumerate}
\item[(a)] Given the imputed complete data $D_n^*$, draw $\kappa^*$ from $p(\kappa\mid  D_{n}^*)$.
\item[(b)] Given the imputed complete data $D_n^*$ and the sampled model $\kappa^*$, draw $\theta^*$ from $p(\theta \mid D_{n}^*,\kappa^*)$.
\item[(c)] Given the model $\kappa^*$ and the parameter $\theta^*$, generate $y_i^*$ from $f(y_i\mid  x_i;\theta^*,\kappa^*)$ independently for each missing unit $i$ and then compute the complete data $D_n^{*}=\{X_n,Y_n^{*}\}$ using $Y^*_n=\{Y_\obs,Y_\mis^*\}$.
\end{enumerate}

By the ergodic theorem \citep{Birkhoff:1931}, as the number of iterations goes to infinity, the limiting distribution of $(Y^*_{\mis},\kappa^*)$ in the above data augmentation algorithm is $p(Y_{\mis},\kappa\mid D_{\obs})=p(Y_{\mis}\mid D_{\obs},\kappa)p(\kappa\mid D_{\obs})$. In step (a), to generate a model from $p(\kappa\mid D_{n})$, we can use
\begin{eqnarray}\label{model_ABC}
p(\kappa \mid D_{n})&=&\frac{ p(\kappa) \int p(Y_n \mid X_n,\theta_\kappa,\kappa)p(\theta_\kappa \mid \kappa) d\theta_\kappa}{\sum_{\kappa\in \mathcal{K}} p(\kappa) \int p(Y_n \mid X_n,\theta_\kappa,\kappa)p(\theta_\kappa\mid \kappa) d\theta_\kappa},
\end{eqnarray}
where $p(\kappa)$ denotes the prior probability that $\kappa$ is the true model and $\mathcal{K}$ is a set of candidate models under consideration.
Similarly, in step (b), the sample-data likelihood is an essential component of the posterior distribution given model $\kappa$ as follows:
\begin{eqnarray}\label{para_ABC}
p(\theta_{\kappa}   \mid D_{n},\kappa)
&=& \frac{p(Y_n \mid X_n,\theta_k,\kappa)p(\theta_\kappa \mid \kappa)}{\int p(Y_n \mid X_n,\theta_k,\kappa)p(\theta_\kappa \mid \kappa)d \theta_\kappa}.
\end{eqnarray}

If  the sampling process is informative, the sample-data likelihood is not available because $p(Y_{n}\mid X_n ,\theta_\kappa,\kappa) \neq \prod_{i\in A}f(y_i\mid x_i;\theta_\kappa,\kappa)$ under informative sampling. As an alternative way to generate parameters from the posterior distribution under informative sampling, \citet{Kim-Yang:2017} and \citet{Wang:2018} recently proposed to replace the posterior distribution in \eqref{para_ABC} with the following partial posterior distribution:
\begin{eqnarray}\label{part:post}
p_g(\theta_{\kappa}  \mid D_n,\kappa)&=& \frac{g(\hat{\theta}_k \mid \theta_\kappa,\kappa)p(\theta_\kappa \mid \kappa)}{\int g(\hat{\theta}_\kappa \mid \theta_\kappa,\kappa)p(\theta_\kappa \mid \kappa)d\theta_\kappa},
\end{eqnarray}
where $g(\hat{\theta}_\kappa \mid \theta_\kappa,\kappa)$ denotes the sampling distribution of the maximum pseudo likelihood estimator $\hat{\theta}_\kappa$ under model $\kappa$.

\begin{remark}
Under informative sampling, using  \eqref{part:post}, one may consider substituting
\begin{eqnarray*}\label{part:post:M}
p_g(\kappa \mid D_n)&=&\frac{ p(\kappa) \int g(\hat{\theta}_\kappa \mid \theta_\kappa,\kappa)p(\theta_\kappa \mid \kappa) d\theta_\kappa}{\sum_{\kappa\in \mathcal{K}} p(\kappa) \int g(\hat{\theta}_\kappa \mid \theta_\kappa,\kappa)p(\theta_\kappa \mid \kappa) d\theta_\kappa }
\end{eqnarray*}
for the posterior model distribution in \eqref{model_ABC}. However, under non-informative (flat or diffuse) priors, this partial posterior model distribution suffers from an identifiability issue which is discussed in Appendix \ref{App:4}.
\end{remark}

To avoid the identifiability issue, we propose to employ the sampling distribution of the unconstrained pseudo likelihood estimator $\hat{\theta}$, where $\hat{\theta}$ is obtained by maximizing \eqref{pMLE} without any restrictions on $\theta$ and use
\begin{eqnarray}\label{part:post:M2}
p_{g}(\kappa \mid D_n)&=&\frac{ p(\kappa) \int g(\hat{\theta}\mid \theta_\kappa,\kappa) p(\theta_\kappa\mid \kappa) d\theta_\kappa}{\sum_{\kappa\in \mathcal{K}} p(\kappa) \int g (\hat{\theta}\mid \theta_\kappa,\kappa) p(\theta_\kappa\mid \kappa) d\theta_\kappa },
\end{eqnarray}
where $g(\hat{\theta} \mid \theta_\kappa,\kappa)$ is  the sampling distribution of $\hat{\theta}$ under model $\kappa$.

We thus modify  steps (a) -- (b) as follows:
\begin{enumerate}
\item[(a$^\star$)] Given the imputed complete data $D_n^*$, draw $\kappa^*$ from $p_{g}(\kappa \mid D_{n}^*)$ in \eqref{part:post:M2}.
\item[(b$^\star$)] Given the imputed  data $D_n^*$ and the sampled model $\kappa^*$, draw $\theta_{\kappa^*}^*$ from $p_{g}(\theta_\kappa\mid D_n^*,\kappa^*)$ in \eqref{part:post}.
\end{enumerate}

\begin{remark}
If equation \eqref{part:post:M2} is available in a closed form, we can easily generate $\kappa^*$ from $p_{g}(\kappa \mid D_{n})$ in step (a$^\star$) by drawing a random sample from a categorical distribution with probabilities $p_{g}(\kappa \mid D_{n})$, $\kappa\in \mathcal{K}$. In general, however, the integrals in \eqref{part:post:M2} are intractable analytically. In this case, numerical approaches such as Laplace's method \citep{Tierney:1986, Tierney:1989} can be used  for approximation. In Theorem \ref{thm:sec4.1} of Section \ref{main:result}, we derive a useful approximation for $p_{g}(\kappa \mid D_{n})$ which facilitates the categorical distribution sampling in step (a$^\star$).
\end{remark}

To sum up, our strategy for implementing the new multiple imputation procedure is as follows: (Model sampling and Imputation) We first generate $M$ samples of missing dataset and model, $(Y_{\mis}^{(1)},\kappa^{(1)}),\ldots,(Y_{\mis}^{(M)},\kappa^{(M)})$, through the new data augmentation steps, (a$^\star$), (b$^\star$) and (c).  We then implement the remaining two steps 3$^\star$ and 4$^\star$ to obtain the proposed multiple imputation estimator and its variance estimator in \eqref{new:rubin}.

\section{Main results}\label{main:result}
In this section, we establish asymptotic properties of the proposed method. Our results rely on the following regularity assumptions that are analogous to \citet{Kim-Yang:2017}:
\begin{assumption}\label{A1}
The regularity conditions for the asymptotic normality of the pseudo maximum likelihood estimator hold for the sequence of finite populations and samples under the unconstrained model.
\end{assumption}
\begin{assumption}\label{A2} Let $\Theta_\kappa$ be the (active) parameter space under model $\kappa$. The prior distribution under model $\kappa$ satisfies the Lipschitz condition over $\Theta_\kappa$, that is, there exists
a positive constant $c_\kappa(<\infty)$ such that $|p(\theta_{\kappa}|\kappa)-p(\vartheta_{\kappa} |\kappa)|\leq c_\kappa\|\theta_{\kappa}-\vartheta_{\kappa}\|$ for $\theta_{\kappa},\vartheta_{\kappa}\in \Theta_\kappa$.
\end{assumption}
\begin{assumption} \label{A3}
Let $\mathcal{B}_n$ be an open ball with center $\theta^\dagger$ and radius $r_n=O(n^{\rho-1/2})$ for $0<\rho<1/2$, where $\theta^\dagger$ indicates the true parameter under the unconstrained model. For any $\theta \in \mathcal{B}_n$, the variance estimator $\hat{V}(\hat{\theta})$ satisfies $\var(\hat{\theta}|\theta)=\hat{V}(\hat{\theta})\{1+o_p(1)\}$ and $(\theta-\hat{\theta})^\T \var(\hat{\theta}|\theta)^{-1}(\theta-\hat{\theta})=(\theta-\hat{\theta})^\T\hat{V}(\hat{\theta})^{-1}(\theta-\hat{\theta})\{1+o_p(1)\}$ as $n\to \infty$.
\end{assumption}
Let $\hat{\theta}$ and $\hat{V}(\hat{\theta})$ be the pseudo maximum likelihood estimator and the variance estimator under the unconstrained model. We can decompose $\hat{\theta}$ and $\hat{V}(\hat{\theta})$ into
$$\hat{\theta}=\left(\begin{matrix} \tilde{\theta}_{\kappa}\\ \tilde{\theta}_{\bar{\kappa}} \end{matrix}\right),\quad \hat{V}(\hat{\theta})=\left(\begin{matrix} \tilde{V}_{\kappa,\kappa} & \tilde{V}_{\kappa,\bar{\kappa}} \\ \tilde{V}_{\bar{\kappa},\kappa}&\tilde{V}_{\bar{\kappa},\bar{\kappa}} \end{matrix}\right),$$
corresponding to $\theta=(\theta^\T_\kappa,\theta_{\bar{\kappa}}^\T)^\T$.
 Define $\tilde{\theta}_\kappa^0=\tilde{\theta}_\kappa+\tilde{V}_{\kappa,\bar{\kappa}} (\tilde{V}_{\bar{\kappa},\bar{\kappa}})^{-1}(\theta^0_{\bar{\kappa}}-\tilde{\theta}_{\bar{\kappa}})$, where $\theta^0_{\bar{\kappa}}$ is the constrained parameter under model $\kappa$. First, we study an asymptotic behavior of the proposed posterior model distribution.
\begin{theorem}\label{thm:sec4.1} Let $p_{g}(\kappa \mid D_n) $ be the partial posterior distribution of model $\kappa$ defined in \eqref{part:post:M2}. Define
$$\tilde{p}_{g}(D_n\mid \kappa) =\begin{cases} \phi( {\theta}_{\bar{\kappa}}^0\mid  \tilde{\theta}_{\bar{\kappa}}, \tilde{V}_{\bar{\kappa},\bar{\kappa}} )p (\tilde{\theta}_\kappa^0 \mid \kappa),\quad& \text{if $|\bar{k}|>0$} \\
p (\tilde{\theta}_\kappa \mid \kappa) &\text{if $|\bar{k}|=0$} \end{cases} ,$$
where $\phi(\cdot\mid \mu,\Sigma)$ is the probability density function of $\mathcal{N}(\mu,\Sigma)$ and $p(\theta_\kappa\mid \kappa)$ is the prior distribution of $\theta_\kappa$ under model $\kappa$. Under Assumptions \ref{A1} -- \ref{A3}, conditional on the complete data $D_n$,
\begin{eqnarray*}
 \left| p_{g}(\kappa \mid D_n) - \frac{\tilde{p}_{g}(D_n\mid \kappa) p(\kappa)}{\sum_{\kappa \in\mathcal{K}}\tilde{p}_{g}(D_n\mid \kappa) p(\kappa)} \right| \to 0,
\end{eqnarray*}
in probability as $n\to \infty$ for $\kappa \in \mathcal{K}$.
\end{theorem}
The proof of Theorem \ref{thm:sec4.1} is shown in Appendix \ref{App:1}. Theorem \ref{thm:sec4.1} leads to a useful approximation to $p_{g}(\kappa \mid D_n)$, which can be used for the proposed step (a$^\star$) in Section \ref{sec:3}.

The Bayesian model selection consistency assures that the posterior probability of the most parsimonious true model tends to one as the sample size goes to infinity \citep{Casella:2009}. The following theorem shows that the proposed method achieves  the model selection consistency.
\begin{theorem}\label{thm:sec4.2} Let $\tau$ be the most parsimonious true model. Suppose that $p(\tau)>0$ regardless of $n$. Under Assumptions \ref{A1} -- \ref{A3}, conditional on the complete data $D_n$,
$$p_{g}(\tau \mid D_n)\to 1 $$
in probability as $n\to \infty$, if $\sqrt{n} (\hat{\theta}-\theta^\dagger)$ is bounded in probability and
$$ 0<\frac{p(\tilde{\theta}_{\kappa_1}^0|{\kappa_1})}{p(\tilde{\theta}_{\kappa_2}^0|{\kappa_2})}<\infty$$
as $n\to \infty$ for any ${\kappa_1},{\kappa_2}\in\mathcal{K}$.
\end{theorem}
The proof of Theorem \ref{thm:sec4.2} is shown in Appendix \ref{App:2}. Under our missing data setup, it is also important to show that $p_{g}(\tau \mid D_\obs)\to 1 $ as $n\to \infty$. Applying Theorem \ref{thm:sec4.2} and noting that $p_{g}(D_\obs\mid \tau)= \int p_{g}(D_{n}\mid \tau) d Y_\mis$, as $n\to \infty$, we have
\begin{eqnarray*}
p_{g}(D_\obs\mid \tau)=\int \frac{p_{g}( \tau\mid D_n)p_{g}(D_n)}{p(\tau)} d Y_\mis
\to \int  p_{g}(D_n) / p(\tau) d Y_\mis=p_{g}(D_\obs)/p(\tau ),
\end{eqnarray*}
in probability, where $p_{g}(D_n\mid \tau)= \int g(\hat{\theta}\mid \theta_\tau,\tau) p(\theta_\tau \mid \tau) d\theta_\tau $. By Bayes' theorem, this thus implies that $p_{g}(\tau \mid D_\obs)=p_{g}(D_\obs \mid \tau )p(\tau )/p_{g}(D_\obs)\to 1$ in probability as $n\to \infty$.

The next theorem shows that the proposed multiple imputation method is asymptotically equivalent to the posterior inference given the observed data under  the true model when the sample size is sufficiently large.
\begin{theorem}\label{thm:sec4.3} Given the most parsimonious true model $\tau$ and the observed data $D_\obs$, define
$$p_{g}(\theta_{\tau}|D_\obs,\tau)= \frac{\int g(\hat{\theta}_\tau \mid \theta_\tau,\tau)p(\theta_\tau \mid \tau)dY_{\mis}}{\int \int g(\hat{\theta}_\tau \mid \theta_\tau,\tau)p(\theta_\tau \mid \tau)dY_{\mis}d\theta_\tau}.$$
Under the sufficient conditions in Theorem \ref{thm:sec4.2} and the population missingness at random assumption, we have
$$p\lim_{M\to \infty} \hat{\theta}_{\MI}=\left[\begin{matrix} E_{g}(\theta_\tau \mid D_\obs,\tau )  \\ \theta^\dagger_{\bar{\tau}} \end{matrix} \right] ,\quad
p\lim_{M\to \infty}  \hat{V}_{\MI}=\left[\begin{matrix} \var_{g}(\theta_\tau \mid D_\obs, \tau) &0_{|\tau|\times |\bar{\tau}|} \\0_{|\bar{\tau}|\times |\tau|}& 0_{|\bar{\tau}|\times |\bar{\tau}|} \end{matrix} \right],$$
in probability as $n\to \infty$, where the conditional expectations are with respect to $p_{g}(\theta_{\tau}|D_\obs,\tau)$ and $\theta^\dagger_{\bar{\tau}}$ is the sub-vector of $\theta^\dagger$ corresponding to $\theta_{\bar{\tau}}$.
\end{theorem}
The proof of Theorem \ref{thm:sec4.3} is given in Appendix \ref{App:3}. Theorem \ref{thm:sec4.3} is similar to Theorem 1 of \citet{Kim-Yang:2017}. According to Lemma 1 of \citet{Kim-Yang:2017}, Theorem \ref{thm:sec4.3} implies that $(\hat{V}^\MI_{\tau,\tau})^{-1/2}(\hat{\theta}^\MI_{\tau}-\theta_{\tau}^\dagger)\to \mathcal{N}(0,I)$ in distribution and $\hat{\theta}_{\bar{\tau}}^\MI-\theta^\dagger_{\bar{\tau}}=o_p(1)$ for sufficiently large $n$ and $M$, where $\hat{\theta}^\MI_{\tau}$ and $\hat{\theta}_{\bar{\tau}}^\MI$ are sub-vectors of $\hat{\theta}_{\MI}$ corresponding to $\theta_{\tau}$ and $\theta_{\bar{\tau}}$, respectively, and $\hat{V}^\MI_{\tau,\tau}$ is a sub-matrix of $\hat{V}_\MI$ corresponding to $\hat{\theta}^\MI_{\tau}$.

\section{Simulation study}
To assess the finite-sample performance of the proposed method, we conduct a limited simulation study for variable selection, which is one of the most important issues in regression analysis. We consider two scenarios under informative sampling design.

In Scenario I, we generate continuous response data as follows: First, we generate a finite population of size $N=30,000$ from the following superpopulation model: $y_i=\beta_0+\beta_1x_{i1}+\cdots+\beta_{12}x_{i12}+\epsilon_i$, where $x_{ij}\overset{iid}{\sim}\mathcal{N}(2,2)$, $\epsilon_i\overset{iid}{\sim}\mathcal{N}(0,\sigma^2)$, and $\theta=(\beta_0,\beta_1,\beta_2,\ldots,{\beta_{12}},\sigma^2)=(-0.5 ,1, 0, \ldots, 0 ,1)$. Second, we generate the response indicator $\delta_i$ from $\text{Bernoulli}(\psi_i)$ independently, where $\text{logit}(\psi_i)=0.2+0.1x_{1i}$. Third, we select a sample from the finite population using Poisson sampling with the sampling indicator $I_i \overset{ind}{\sim} \text{Bernoulli}(\pi_i)$, where $\text{logit}(1-\pi_i)=4.5-0.2y_i$. In this scenario, the true data-generating model $\tau$ is $y_i\overset{ind}{\sim} \mathcal{N}(\mu_{\tau i},\sigma^2)$ with $\mu_{\tau i}=\beta_0+\beta_1x_{i1}$. In Scenario II, we generate binary response data as follows: First, we generate a finite population of size $N=30,000$ from the following superpopulation model: $y_i\overset{ind}{\sim}\text{Bernoulli}(\mu_i)$ with $\text{logit}(\mu_i)=\beta_0+\beta_1x_{i1}+\cdots+\beta_{12}x_{i12}$, where $x_{ij}\overset{iid}{\sim}\mathcal{N}(1,2)$ and $\theta=(\beta_0,\beta_1,\beta_2,\ldots,{\beta_{12}})=(-0.5,1, 0, \ldots , {0})$. Second, we generate the response indicator $\delta_i$ from $\text{Bernoulli}(\psi_i)$ independently, where $\text{logit}(\psi_i)=0.2+0.2x_{1i}$. Third, we select a sample from the finite population using Poisson sampling with the sampling indicator $I_i \overset{ind}{\sim} \text{Bernoulli}(\pi_i)$, where $\text{logit}(1-\pi_i)=4.4-0.3 y_i$. In Scenario II, the true data-generating model $\tau$ is $y_i\overset{ind}{\sim} \text{Bernoulli}(\mu_{\tau i})$ with $\text{logit}(\mu_{\tau i})=\beta_0+\beta_1x_{i1}$. For each scenario, we repeat $3,000$ Monte Carlo simulations. For both scenarios, the sampling mechanism is non-ignorable. In addition, since $y_i$ and $\delta_i$ are conditionally independent given $x_i$, the missingness at random holds at population level. The average response rate is around $60\%$ for both scenarios. The sample sizes range from $397$ to $541$ with median $470$ in Scenario I and from $371$ to $514$ with median $438$ in Scenario II.

The parameters of interest are the relevant coefficients of the true data-generating model, $(\beta_0,\beta_1)=(-0.5,1)$. For the candidate model set $\mathcal{K}$, we consider all possible $2^{12}(=4,096)$ regression models, where every model in $\mathcal{K}$ includes the intercept. We assume that there is no preferred model in $\mathcal{K}$ and we thus define $p(\kappa)\propto 1$ for $\kappa \in \mathcal{K}$. Let $\mathcal{A}_{\kappa}$ be an index set of active coefficients under model $\kappa$. For each model $\kappa$, to make the prior distribution of the parameter relatively flat, we assign $\beta_j \overset{iid}{\sim}\mathcal{N}(0,10^5)$ for $j\in \mathcal{A}_\kappa$ and $p(\beta_j=0)=1$ for $j\notin \mathcal{A}_\kappa$ in both scenarios, and $\log\sigma^2\sim \mathcal{N}(0,10^5)$ in Scenario I. It is straightforward to check that our priors satisfy the sufficient conditions to achieve the model selection consistency in Theorem \ref{thm:sec4.2}.

We compare the following four multiple imputation methods based on imputation size $M=100$: (i) multiple imputation under the true model, which is a benchmark for comparison; (ii) multiple imputation under the full model; (iii) multiple imputation under the best model selected by LASSO \citep{Tibshirani:1996}; (iv) the proposed multiple imputation. In (i), under the true model, we implement the multiple imputation procedure of \citet{Kim-Yang:2017} (say KY-MI), which is most suitable for non-ignorable sampling mechanism. In (ii), KY-MI is used under the full model. In (iii), assuming the complete data is available, we first select a single best model by maximizing the pseudo log-likelihood of \eqref{pMLE} subject to the LASSO penalty, where the regularization parameter is chosen by 10-fold cross-validation. Then, given the best model, KY-MI is applied to the observed data. In (iv), we employ the proposed method with the asymptotic distributions of $p_{g}(\kappa \mid  D_n)$ and $p_{g}(\theta_\kappa\mid D_n,\kappa)$ derived in Section \ref{main:result}. For the sampling weights, we use $w_i=1/\pi_i$.

The simulation results are summarized in Table \ref{table:1}. In both scenarios, the performance of the proposed method ($\MI_\BMA$) is always comparable to that of the imputation method based on the true model ($\MI_\TRUE$). Furthermore, the true positive rate (TPR) and the true negative rate (TNR) of $\MI_\BMA$ are close to $100(\%)$ in both scenarios. Indeed, the results of our simulation study agree with our theoretical results in Section \ref{main:result}. However, as the full model includes all the irreverent variables, the full model-based approach ($\MI_{\FULL}$) has large variance and mean squared error (MSE), especially for the intercept ($\beta_0$) estimate. Similarly, the model selection approach ($\MI_{\LASSO}$) involves large variance and mean squared error for the intercept estimate. The true negative rate of variable selection using the LASSO is low, less than $80\%$ in both scenarios, and this means that the LASSO method tends to select an overfitted model. Furthermore, the model selection approach provides poor coverage probability (CP) of $95\%$ confidence interval for the intercept estimate. This is due to the fact that the best model-based method ignores the uncertainty associated with variable selection. Our results clearly show that implementing the multiple imputation procedures under a single best model leads to underestimation of the variance of multiple imputation estimator.

 \begin{table}[ht]
 \caption{Simulation results based on Monte Carlo sample  size $=3,000$. Notation: CP --- Coverage Probability; Var --- Variance; MSE --- Mean Squared Error; TPR --- True Positive Rate; TNR --- True Negative Rate.}
\label{table:1}\par
\vskip .2cm
\centerline{\tabcolsep=3truept\begin{tabular}{|c|c|c|c|c|c|c|c|c|c|c|c|}
\hline
Scenario& Method & \multicolumn{2}{c|}{CP} & \multicolumn{2}{c|}{Var} & \multicolumn{2}{c|}{Bias} & \multicolumn{2}{c|}{MSE} &TPR &TNR  \\
&  & \multicolumn{2}{c|}{($95\%$)} & \multicolumn{2}{c|}{($\times 10^{4}$)} & \multicolumn{2}{c|}{($\times 10$)} & \multicolumn{2}{c|}{($\times 10^{4}$)} &(\%) &(\%) \\
& & $\beta_0$ & $\beta_1$ & $\beta_0$ &$\beta_1$ & $\beta_0$ &$\beta_1$ & $\beta_0$ &$\beta_1$ & & \\
\hline
I&$\MI_{\TRUE}$& 94.6 & 95.0 & 169.6 & 22.4 &0.0 &0.0 & 174.0 & 22.9 & 100 & 100 \\
&$\MI_{\BMA}$& 94.6 & 95.2 & 175.9 & 22.4 &0.0 &0.0 &178.1 &22.8 & 100 & 99.9  \\
&$\MI_{\LASSO}$& 88.0 & 95.0 & 352.6 & 21.9 &0.1 &0.0 & 592.1  & 23.0 &100 & 78.5 \\
&$\MI_{\FULL}$&93.5 &93.9 & 1045.0 & 21.7 &0.1 &0.0 & 1133.2 & 23.5 & 100& 0 \\
\hline
II&$\MI_{\TRUE}$& 95.5 & 94.7 & 402.9 & 220.5&0.0 &0.1& 391.0 & 230.2 & 100&100 \\
&$\MI_{\BMA}$& 95.4 & 94.6 & 409.9 & 220.7&0.0 &0.1 & 394.5 & 230.9 &100 &99.9 \\
&$\MI_{\LASSO}$& 89.9 & 94.8 & 754.0 & 233.0&0.0 &0.4 & 1157.5 & 266.1 &100 &76.5 \\
&$\MI_{\FULL}$&93.9 & 93.9 & 1913.6 & 253.5&0.0&0.7 & 2139.6 & 331.5 &100 &0 \\
\hline
\end{tabular}}
\end{table}

\section{Concluding remarks}

We have developed a new multiple imputation method using the idea of Bayesian model averaging. The proposed method provides valid inference incorporating the uncertainty of model selection and also often achieves more efficient estimator than the method based on  a single best model. For example, in our simulation study in Section 5, our proposed method shows smaller MSE than the multiple imputation estimator  using Lasso-based model selection. The Bayesian approach captures all the uncertainly automatically and the computation is relatively simple.

The proposed method can be directly extended to handle high dimensional model problem for imputation. Extensions to tree-based methods, such as random forests \citep{Breiman:2001}, can also be interesting. Such extensions will be topics for future research.

\appendix
\section*{Appendix}
This appendix includes the proofs of Theorems 1, 2 and 3 and a discussion on the identifiability issue. In our proofs, Lemma 1 of \citet{Kim-Yang:2017} is needed. For the sake of readability, we restate the lemma as follows.
\begin{lemma}\label{lemma:A1} Under Assumptions 1, 2 and 3, conditional on the full sample data and model $\kappa (\supset \tau)$,
$$\int \left| p_g(\theta_\kappa\mid D_n,\kappa)- \phi(\theta_{\kappa}\mid \hat{\theta}_\kappa,\hat{V}(\hat{\theta}_\kappa))\right|d\theta_k \to 0$$
in probability, as $n\to \infty$.
\end{lemma}
\section{Proof of Lemma 1} In the proof of Lemma 1,  \citet{Kim-Yang:2017} showed that as $n\to \infty$,
\begin{eqnarray}\label{eq:l1}
p_g(\theta_\kappa\mid D_n,\kappa)-\phi(\theta_{\kappa}\mid \hat{\theta}_\kappa,\hat{V}(\hat{\theta}_\kappa))=o_p(1),
\end{eqnarray}
for $\theta_\kappa\in \mathcal{B}_{\kappa,n}$, where $\mathcal{B}_{\kappa,n}$ denotes an open ball with center $\theta^\dagger_\kappa$ and radius $r_n=O(n^{\rho-1/2})$ for $0<\rho<1/2$. By the reverse Fatou's lemma, \eqref{eq:l1} implies
\begin{eqnarray}\label{eq:part1}
p\lim_{n\to \infty} \int_{\mathcal{B}_{\kappa,n}} \left| p_g(\theta_\kappa\mid D_n,\kappa)- \phi(\theta_{\kappa}\mid \hat{\theta}_\kappa,\hat{V}(\hat{\theta}_\kappa))\right|d\theta_{\kappa}=0.
\end{eqnarray}
In the proof of Lemma 1,  \citet{Kim-Yang:2017} also showed that
\begin{eqnarray}\label{eq:l2}
p\lim_{n\to \infty} \int_{\mathcal{B}_{\kappa,n}} \phi(\theta_{\kappa}\mid \hat{\theta}_\kappa,\hat{V}(\hat{\theta}_\kappa)) d\theta_{\kappa}=1.
\end{eqnarray}
From \eqref{eq:l1} and \eqref{eq:l2}, we have that
\begin{eqnarray}
\nonumber &&\int_{\mathcal{B}_{\kappa,n}^c} \left| p_g(\theta_\kappa\mid D_n,\kappa)- \phi(\theta_{\kappa}\mid \hat{\theta}_\kappa,\hat{V}(\hat{\theta}_\kappa))\right|d\theta_{\kappa}\\
\nonumber&&\leq\int_{\mathcal{B}_{\kappa,n}^c} p_g(\theta_\kappa\mid D_n,\kappa)d\theta_{\kappa}+ \int_{\mathcal{B}_{\kappa,n}^c} \phi(\theta_{\kappa}\mid \hat{\theta}_\kappa,\hat{V}(\hat{\theta}_\kappa))d\theta_{\kappa}\\
\nonumber&&=1-\int_{\mathcal{B}_{\kappa,n}} p_g(\theta_\kappa\mid D_n,\kappa)d\theta_{\kappa}+1- \int_{\mathcal{B}_{\kappa,n}} \phi(\theta_{\kappa}\mid \hat{\theta}_\kappa,\hat{V}(\hat{\theta}_\kappa))d\theta_{\kappa}\\
 \label{eq:part2} &&\to 0
\end{eqnarray}
in probability as $n\to \infty$. Hence, \eqref{eq:part1} and  \eqref{eq:part2} immediately complete our proof.

\section{Proof of Theorem 1}\label{App:1}
Note that
$$p_{g}(\kappa \mid D_n) =\frac{{p}_{g}(D_n\mid \kappa) p(\kappa)}{\sum_{\kappa \in\mathcal{K}}{p}_{g}(D_n\mid \kappa) p(\kappa)},$$
where ${p}_{g}(D_n\mid \kappa)=\int g(\hat{\theta} \mid \theta_{\kappa},\kappa) p(\theta_\kappa\mid \kappa) d\theta_\kappa$. Hence it suffices to show that $$p_{g}(D_n\mid \kappa) =\begin{cases} \phi( {\theta}_{\bar{\kappa}}^0\mid  \tilde{\theta}_{\bar{\kappa}}, \tilde{V}_{\bar{\kappa},\bar{\kappa}} )p (\tilde{\theta}_\kappa^0 \mid \kappa),\quad& \text{if $|\bar{k}|>0$} \\
p (\tilde{\theta}_\kappa\mid \kappa), &\text{if $|\bar{k}|=0$} \end{cases} $$ in probability as $n\to \infty$.

First, we assume $|\bar{\kappa}|>0$, that is, $\kappa$ is a restricted model. By Assumptions 1 and 3, we have
\begin{eqnarray}\label{thm1:eq:1}
g(\hat{\theta}\mid \theta) =\phi(\hat{\theta} \mid \theta , \hat{V}(\hat{\theta}))\{1+o_p(1)\}.
\end{eqnarray}
It is straightforward to check that the multivariate normal density function can be factored as
\begin{eqnarray}\label{thm1:eq:2}
\phi(\hat{\theta} \mid \theta , \hat{V}(\hat{\theta}) )=  \phi( \theta_\kappa \mid \tilde{\theta}_{\kappa}+\tilde{V}_{{\kappa},\bar{\kappa}}\tilde{V}_{\bar{\kappa},\bar{\kappa}}^{-1} ( \theta_{\bar{\kappa}}-
\tilde{\theta}_{\bar{\kappa}}),\tilde{V}_{\kappa,\kappa}-\tilde{V}_{{\kappa},\bar{\kappa}}\tilde{V}_{\bar{\kappa},\bar{\kappa}}^{-1}\tilde{V}_{\bar{\kappa},{\kappa}} )  \phi( \theta_{\bar{\kappa}}\mid  \tilde{\theta}_{\bar{\kappa}} ,  \tilde{V}_{\bar{\kappa},\bar {\kappa}} ).
\end{eqnarray}
Recall that, under model $\kappa$, we impose the restriction $\theta_{\bar{\kappa}}=\theta_{\bar{\kappa}}^0$ on $\theta$. Following from \eqref{thm1:eq:1} and \eqref{thm1:eq:2}, the partial likelihood under model $\kappa$ is
\begin{eqnarray}\label{thm4:eq1}
 g(\hat{\theta} \mid \theta_{\kappa},\kappa) = \phi( \theta_\kappa \mid \tilde{\theta}_\kappa^0 ,\tilde{V}_{\kappa,\kappa}-\tilde{V}_{{\kappa},\bar{\kappa}}\tilde{V}_{\bar{\kappa},\bar{\kappa}}^{-1}\tilde{V}_{\bar{\kappa},{\kappa}} )  \phi( \theta_{\bar{\kappa}}^0\mid  \tilde{\theta}_{\bar{\kappa}} ,  \tilde{V}_{\bar{\kappa},\bar {\kappa}} )\{1+o_p(1)\},
\end{eqnarray}
where $\tilde{\theta}_\kappa^0= \tilde{\theta}_{\kappa}+\tilde{V}_{{\kappa},\bar{\kappa}}\tilde{V}_{\bar{\kappa}, \bar{\kappa}}^{-1} ( \theta_{\bar{\kappa}}^0-
\tilde{\theta}_{\bar{\kappa}})$. To calculate the partial marginal likelihood $p_{g}(D_n|\kappa)$, we apply the formula of Laplace's method \citep{Tierney:1986, Tierney:1989},
$$\int b(\theta_\kappa)\exp\{-nh(\theta_\kappa)\} d\theta_\kappa =\left(\frac{2\pi }{n}\right)^{|\kappa|/2}|\tilde{\Sigma}_\kappa|^{1/2} b(\tilde{\theta}_\kappa)\exp\{-nh(\tilde{\theta}_\kappa)\} \{1+O(n^{-1})\},  $$
as $n\to \infty$, where $\tilde{\theta}_\kappa$ is the minimizer of $h(\cdot)$ and $\tilde{\Sigma}_\kappa$ is the inverse of the Hessian matrix of $h$, evaluated at $\tilde{\theta}_\kappa$. By taking $h(\theta_\kappa)=-\frac{1}{n} \log  \phi( \theta_\kappa \mid \tilde{\theta}_\kappa^0 ,\tilde{V}_{\kappa,\kappa}-\tilde{V}_{{\kappa},\bar{\kappa}}\tilde{V}_{\bar{\kappa},\bar{\kappa}}^{-1}\tilde{V}_{\bar{\kappa},{\kappa}} )$ and $b(\theta_\kappa)=\pi(\theta_\kappa\mid \kappa)$, as $n\to \infty$, the Laplace's method yields
\begin{eqnarray*}
p_{g}(D_n|\kappa)&=&\int g(\hat{\theta} \mid \theta_{\kappa},\kappa) p(\theta_\kappa\mid \kappa) d\theta_\kappa \\
&=& \phi( \theta_{\bar{\kappa}}^0\mid  \tilde{\theta}_{\bar{\kappa}} ,  \tilde{V}_{\bar{\kappa},\bar {\kappa}} ) \left\{ \int \phi( \theta_\kappa \mid \tilde{\theta}_\kappa^0 ,\tilde{V}_{\kappa,\kappa}-\tilde{V}_{{\kappa},\bar{\kappa}}\tilde{V}_{\bar{\kappa},\bar{\kappa}}^{-1}\tilde{V}_{\bar{\kappa},{\kappa}} ) p(\theta_\kappa\mid \kappa) d\theta_\kappa \right\}\{1+o_p(1)\}\\
&=& \phi( \theta_{\bar{\kappa}}^0\mid  \tilde{\theta}_{\bar{\kappa}} ,  \tilde{V}_{\bar{\kappa},\bar {\kappa}} )  p(\tilde{\theta}^0_\kappa\mid \kappa)\{1+o_p(1)\},
\end{eqnarray*}
which completes the first part of our proof. Second, we assume $|\bar{\kappa}|=0$, that is, $\kappa$ is the unconstrained model. Note that, under the unconstrained model $\kappa$, $g(\hat{\theta}\mid \theta_\kappa,\kappa)=g(\hat{\theta}\mid \theta)$ and $\hat{\theta}=\tilde{\theta}_\kappa$ since $\theta_\kappa=\theta$. From Lemma \ref{lemma:A1} and \eqref{thm1:eq:1},
$$p_{g}(D_n\mid \kappa)=\frac{g(\hat{\theta}\mid \tilde{\theta}_\kappa,\kappa)p( \tilde{\theta}_\kappa\mid \kappa )}{p_{g}( \tilde{\theta}_\kappa \mid D_n,\kappa)}=\frac{g(\hat{\theta}\mid\hat{ \theta}) p( \tilde{\theta}_\kappa\mid \kappa )}{\phi(\hat{\theta} \mid \hat{\theta} , \hat{V}(\hat{\theta}))}=p( \tilde{\theta}_\kappa\mid \kappa )$$
in probability as $n\to\infty$. This completes the proof of Theorem 1.

\section{Proof of Theorem 2}\label{App:2}
Let $\mathcal{K}_1=\mathcal{K}\setminus \{\tau\}$ be a set of candidate models excluding the most parsimonious true model. Since $\sum_{\kappa\in \mathcal{K}}p_{g}(\kappa\mid D_n)=1$, to complete the proof of Theorem 2, it suffices to show that $p_{g}(\kappa\mid D_n)=0$ in probability as $n\to \infty$ for any $\kappa \in\mathcal{K}_1$. From Theorem 1, as $n\to \infty$, we have
$$p_{g}(\kappa\mid D_n)= \frac{\tilde{p}_{g}(D_n\mid \kappa)p(\kappa) }{\sum_{\kappa\in \mathcal{K}}\tilde{p}_{g}(D_n\mid \kappa) p(\kappa)}\leq \frac{\tilde{p}_{g}(D_n\mid \kappa)p(\kappa) }{\tilde{p}_{g}(D_n\mid \gamma) p(\gamma)}$$
in probability for any model $\gamma\in \mathcal{K}$. Hence, the proof can be done by showing that there always exists a model $\kappa^\star\in \mathcal{K}$ such that
 $\tilde{p}_{g}(D_n\mid \kappa)p(\kappa) /\{\tilde{p}_{g}(D_n\mid\kappa^\star) p(\kappa^\star)\}\to 0$ in probability as $n\to \infty$ for any $\kappa \in \mathcal{K}_1$. To this end, we partition $\mathcal{K}_1$ into two sets, $\mathcal{K}_2=\{\kappa\in \mathcal{K}_1 :\tau \not\subset \kappa\}$ (misspecified models) and $\mathcal{K}_3=\{\kappa\in \mathcal{K}_1  :\tau \subset \kappa \}$ (overfitted models). First we consider the case that $\kappa \in \mathcal{K}_2$, that is, $\kappa$ is a misspecified model. Let $\theta_{\bar{\kappa}}^\dagger$, $\tilde{\theta}_{\bar{\kappa}}$ and $\tilde{V}_{\bar{\kappa},\bar{\kappa}}$ be, respectively, the sub-vector of $\theta^\dagger$, the sub-vector of $\hat{\theta}$ and the sub-matrix of $\hat{V}(\hat{\theta})$ corresponding to $\theta_{\bar{\kappa}}$. For any $\kappa$ such that $|\bar{\kappa}|>0$, we have
\begin{eqnarray}\label{proof:thm3:eq1}
(\tilde{\theta}_{\bar{\kappa}}-\theta_{\bar{\kappa}}^\dagger)^\T \tilde{V}_{\bar{\kappa},\bar{\kappa}}^{-1}(\tilde{\theta}_{\bar{\kappa}}-\theta_{\bar{\kappa}}^\dagger) \leq \lambda_{\min,{\bar{\kappa}}}^{-1}n\|\tilde{\theta}_{\bar{\kappa}}-\theta_{\bar{\kappa}}^\dagger\|^2,
\end{eqnarray} where $\lambda_{\min,{\bar{\kappa}}}$ is the smallest eigenvalue of $n\tilde{V}_{{\bar{\kappa}},{\bar{\kappa}}}$. We set $\kappa^\star=\kappa \cup \tau$. Since $\kappa^\star$ includes the true model $\tau$, we have $\theta^0_{\bar{\kappa}^\star}=\theta^\dagger_{\bar{\kappa}^\star}$. Then, \eqref{proof:thm3:eq1} implies
\begin{eqnarray}\label{proof:thm3:eq2}
(\tilde{\theta}_{\bar{\kappa}^\star}-\theta_{\bar{\kappa}^\star}^0)^\T \tilde{V}_{\bar{\kappa}^\star,\bar{\kappa}^\star}^{-1}(\tilde{\theta}_{\bar{\kappa}^\star}-\theta_{\bar{\kappa}^\star}^0)  \leq \lambda_{\min,\bar{\kappa}^\star}^{-1}n\|\tilde{\theta}_{{\bar{\kappa}^\star}}-\theta_{{\bar{\kappa}^\star}}^\dagger\|^2.
\end{eqnarray} Define $\xi_0=\inf_{\kappa\in \mathcal{K}_2}\|\theta_{\bar{\kappa}}^0 -\theta_{\bar{\kappa}}^\dagger\|$. Note that $\xi_0>0$ because $\theta_{\bar{\kappa}}^0 \neq \theta_{\bar{\kappa}}^\dagger$ for any $\kappa  \in \mathcal{K}_2$. Since $\sqrt{n} (\hat{\theta}-\theta^\dagger)$ is assumed to be bounded in probability, we have $\tilde{\theta}_{\bar{\kappa}}-{\theta}^\dagger_{\bar{\kappa}}=o_p(1)$. This yields
 \begin{eqnarray}\label{proof:thm3:eq3}
(\tilde{\theta}_{\bar{\kappa}}-\theta_{\bar{\kappa}}^0)^\T \tilde{V}_{\bar{\kappa},\bar{\kappa}}^{-1}(\tilde{\theta}_{\bar{\kappa}}-\theta_{\bar{\kappa}}^0)  \geq \lambda_{\max,\bar{\kappa}}^{-1}{n\|\tilde{\theta}_{\bar{\kappa}}-\theta_{\bar{\kappa}}^0\|^2}\geq \lambda_{\max,\bar{\kappa}}^{-1}{n\xi_0^2}\{1+o_p(1)\}
\end{eqnarray}
for sufficiently large $n$, where $\lambda_{\max,\bar{\kappa}}$ is the largest eigenvalue of $n\tilde{V}_{\bar{\kappa},\bar{\kappa}}$. From \eqref{proof:thm3:eq2} and \eqref{proof:thm3:eq3}, as $n\to \infty$,
$$\frac{\phi(\theta^0_{\bar{\kappa}}\mid \tilde{\theta}_{\bar{\kappa}},\tilde{V}_{\bar{\kappa},\bar{\kappa}})}{ \phi(\theta^0_{\bar{\kappa}^\star}\mid \tilde{\theta}_{\bar{\kappa}^\star},\tilde{V}_{\bar{\kappa}^\star,\bar{\kappa}^\star}) }\leq a_{1} \left(\frac{n}{2\pi}\right)^{(|\kappa^\star|-|\kappa|)/2} \exp\left(-\frac{\xi_0^2}{2\lambda_{\max,\bar{\kappa}} }n+a_2\right)\to 0$$
in probability, where $a_1$ and $a_2$ are positive constants such that $(|n \tilde{V}_{\bar{\kappa}^\star,\bar{\kappa}^\star} |/|n\tilde{V}_{\bar{\kappa},\bar{\kappa}} |)^{1/2}\leq a_1 <\infty$ and $ \lambda_{\min,\bar{\kappa}^\star}^{-1}n\|\tilde{\theta}_{{\bar{\kappa}^\star}}-\theta_{{\bar{\kappa}^\star}}^\dagger\|^2/2\leq a_2 <\infty$. By Theorem 1, if $0<\lim_{n\to\infty} p(\tilde{\theta}_{\kappa}^0\mid {\kappa})/p(\tilde{\theta}_{\kappa^\star}^0\mid {\kappa^\star})<\infty$, then $\tilde{p}_{g}(D_n\mid \kappa)p(\kappa) /\{\tilde{p}_{g}(D_n\mid \kappa^\star) p(\kappa^\star)\}\to 0$ in probability as $n\to \infty$ for any $\kappa \in \mathcal{K}_2$.

Second, we assume $\kappa \in\mathcal{K}_3$, that is, $\kappa$ is an overfitted model. In this case, we set $\kappa^\star=\tau$. Suppose $|\bar{\kappa}|>0$. Following from \eqref{proof:thm3:eq2} and noting that $(\tilde{\theta}_{\bar{\kappa}}-\theta_{\bar{\kappa}}^0)^\T \tilde{V}_{\bar{\kappa},\bar{\kappa}}^{-1}(\tilde{\theta}_{\bar{\kappa}}-\theta_{\bar{\kappa}}^0) \geq 0$, as $n\to \infty$, $$\frac{\phi(\theta^0_{\bar{\kappa}}\mid \tilde{\theta}_{\bar{\kappa}},\tilde{V}_{\bar{\kappa},\bar{\kappa}})}{ \phi(\theta^0_{\bar{\kappa}^\star}\mid \tilde{\theta}_{\bar{\kappa}^\star},\tilde{V}_{\bar{\kappa}^\star,\bar{\kappa}^\star}) }\leq a_{1}  \exp\left(a_2\right)\left(\frac{2\pi}{n}\right)^{(|\kappa|-|\kappa^\star|)/2}\to 0$$
in probability. If  $|\bar{\kappa}|=0$, then
$$\frac{1}{ \phi(\theta^0_{\bar{\kappa}^\star}\mid \tilde{\theta}_{\bar{\kappa}^\star},\tilde{V}_{\bar{\kappa}^\star,\bar{\kappa}^\star}) }\leq |n \tilde{V}_{\bar{\kappa}^\star,\bar{\kappa}^\star} |^{1/2} \exp\left(a_2\right)\left(\frac{2\pi}{n}\right)^{|\bar{\kappa}^\star|/2}\to 0$$
in probability as $n\to \infty$. Then, Theorem 1 leads to $\tilde{p}_{g}(D_n\mid \kappa)p(\kappa) /\{\tilde{p}_{g}(D_n \mid \kappa^\star) p(\kappa^\star)\}\to 0$ in probability as $n\to \infty$ for any $\kappa \in \mathcal{K}_3$. This completes the proof of Theorem 2.

\section{Proof of Theorem 3}\label{App:3}
Recall that $(Y_{\mis}^{(m)},\kappa^{(m)})$ are generated from $p_{g}(Y_{\mis}\mid D_{\obs},\kappa)p_{g}(\kappa\mid D_{\obs})$. In Section 4 of the manuscript, we have shown that Theorem 2 implies $p_{g}(\tau\mid D_{\obs})=1$ in probability as $n\to \infty$. Hence, in the proposed multiple imputation procedure, $Y_{\mis}^{(m)}$ are generated from $p_{g}(Y_{\mis}\mid D_{\obs},\tau)=\int p(Y_{\mis}\mid D_{\obs},\theta_{\tau},\tau)p_{g}(  \theta_{\tau}\mid D_{\obs}, \tau) d\theta_{\tau}$ as $n\to \infty$. Let $\hat{\theta}^{(m)}=(\hat{\theta}^{(m)\T}_{\kappa^{(m)}},\hat{\theta}^{(m)\T}_{\bar{\kappa}^{(m)}})^\T$. Note that Lemma \ref{lemma:A1} implies
$$E_{g}(\theta_\tau\mid D_n,\tau)=\hat{\theta}_\tau \{1+o_p(1)\},\quad \var_{g}(\theta_\tau \mid D_n,\tau )=\hat{V}(\hat{\theta}_\tau)\{1+o_p(1)\}.$$ Hence, we have
\begin{eqnarray*}
p\lim_{M\to \infty} M^{-1} \sum_{m=1}^M \hat{\theta}^{(m)}_{\kappa^{(m)}}&=&p\lim_{M\to \infty} M^{-1} \sum_{m=1}^M \hat{\theta}^{(m)}_{\tau}\\
 &=&\int E_{g}(\theta_\tau\mid D_n,\tau) p_{g}(Y_{\mis}\mid D_{\obs},\tau) d Y_{\mis} \\
&=&E_{g}\{E_{g}(\theta_\tau \mid D_n)\mid D_{\obs},\tau\}\\
&=&E_{g}(\theta_\tau | D_{\obs},\tau)
\end{eqnarray*}
in probability as $n\to \infty$.
This implies that
$$p\lim_{M\to \infty}\hat{\theta}_{\MI}=\left[\begin{matrix}E_g(\theta_{\tau}\mid D_\obs,\tau)\\ ~\\ \theta_{\bar{\tau}}^\dagger \end{matrix}\right].$$
Similarly,
\begin{eqnarray*}
p\lim_{M\to \infty} W&=&\left[\begin{matrix} E_g\{\var_g(\theta_{\tau}\mid D_n,\tau) \mid D_\obs,\tau\}& 0_{|\tau|\times |\bar{\tau}|}\\ 0_{|\bar{\tau}|\times |\tau|}&0_{|\bar{\tau}|\times |\bar{\tau}|} \end{matrix}\right],\\
p\lim_{M\to \infty} B&=&\left[\begin{matrix} \var_g\{E_g(\theta_{\tau}\mid D_n,\tau) \mid D_\obs,\tau \}& 0_{|\tau|\times |\bar{\tau}|}\\ 0_{|\bar{\tau}|\times |\tau|}&0_{|\bar{\tau}|\times |\bar{\tau}|}\end{matrix}\right].
\end{eqnarray*}
Hence, we have
$$p\lim_{M\to \infty}\hat{V}_{\MI}=p\lim_{M\to \infty} W +p\lim_{M\to \infty}B = \left[\begin{matrix} \var_g(\theta_{\tau}\mid D_\obs,\tau) & 0_{|\tau|\times |\bar{\tau}|}\\ 0_{|\bar{\tau}|\times |\tau|}&0_{|\bar{\tau}|\times |\bar{\tau}|}\end{matrix}\right],$$
which completes the proof of Theorem 3.

\section{Identifiability issue}\label{App:4}
\begin{theorem}\label{theorem:E1}
Let $\tau$ and $\kappa$ be, respectively, the true data-generating model and a candidate model such that $\tau \subset \kappa$.  Define
\begin{eqnarray*}
p_g(\kappa \mid D_n)&=&\frac{ p(\kappa) \int g(\hat{\theta}_\kappa \mid \theta_\kappa,\kappa)p(\theta_\kappa \mid \kappa) d\theta_\kappa}{\sum_{\kappa\in \mathcal{K}} p(\kappa) \int g(\hat{\theta}_\kappa \mid \theta_\kappa,\kappa)p(\theta_\kappa \mid \kappa) d\theta_\kappa }.
\end{eqnarray*}
Under Assumptions 1--3 in Section 4 of the manuscript, if $p(\theta_{\kappa}\mid\kappa)p(\kappa)=p(\theta_{\tau}\mid \tau)p(\tau)$, then the true model is non-identifiable in the sense that
$$p_g(\kappa\mid D_n)=p_g(\tau\mid D_n)$$
in probability as $n\to \infty$.
\end{theorem}
The proof of Theorem \ref{theorem:E1} is as follows. Given any model $\kappa (\supset \tau)$, Lemma \ref{lemma:A1} and the asymptotic normality of $\hat{\theta}_\kappa$ imply that $g(\hat{\theta}_\kappa\mid \theta_\kappa,\kappa)=p_g(\theta_\kappa\mid D_n,\kappa)\{1+o_p(1)\}$ for sufficiently large $n$. This yields
$$p_g(D_n\mid \kappa)=\frac{g(\hat{\theta}_\kappa\mid \theta_\kappa,\kappa)p(\theta_\kappa\mid \kappa)}{ p_g(\theta_\kappa\mid D_n,\kappa)}= p(\theta_\kappa\mid \kappa)\{1+o_p(1)\},$$
where $p_g(D_n\mid \kappa)=\int g(\hat{\theta}_\kappa\mid \theta_\kappa,\kappa)p(\theta_\kappa\mid \kappa) d\theta_\kappa$.
By Bayes' theorem, if $p(\theta_{\tau}\mid \tau)p(\tau)=p(\theta_{\kappa}\mid \kappa)p(\kappa)$, then
$${p_g(D_n\mid \kappa) p(\kappa)}={ p(\theta_{\kappa}\mid \kappa)p(\kappa) }={ p(\theta_{\tau}\mid \tau)p(\tau) }={p_g(D_n\mid\tau) p(\tau)}$$
in probability as $n\to \infty$. Hence, for any $\kappa (\supset \tau)$, we have
$$p_g(\kappa \mid D_n)=\frac{p_g(D_n\mid \kappa) p(\kappa)}{p_g(D_n)}=\frac{p_g(D_n\mid \tau) p(\tau)}{p_g(D_n)}=p_g(\tau\mid D_n)$$
in probability as $n\to \infty$, where $p_g(D_n)=\sum_{\kappa\in \mathcal{K}}p_g(D_n\mid \kappa) p(\kappa) $. This completes our proof.

\end{document}